\documentclass[reprint,amsmath,amssymb,aps,showpacs,superscriptaddress,prb]{revtex4-1}

\usepackage{graphicx}
\usepackage{bm}
\usepackage{epsfig}
\usepackage{amssymb}
\usepackage{amsfonts}
\usepackage{braket}
\usepackage{color}
\usepackage{epstopdf}
\usepackage{hyperref}
\usepackage{float}
\restylefloat{table}
\usepackage{bibentry}
\usepackage{multirow}
\usepackage[caption=false]{subfig}
\newcommand{\ba}{\begin{eqnarray}}
\newcommand{\ea}{\end{eqnarray}}
\newcommand{\bd}{\begin{displaymath}}
\renewcommand{\v}[1]{{\bf #1}}
\newcommand{\nn}{\nonumber \\}

\graphicspath{{figures/}}

\begin{document}
\title{Statistical recovery of the classical spin Hamiltonian}

\author{Vinit Kumar Singh}
\email[Electronic address:$~~$]{vinitsingh911@gmail.com}
\affiliation{Department of Physics, Indian Institute of Technology, Kharagpur 721302, India}
\author{Jung Hoon Han}
\email[Electronic address:$~~$]{hanjh@skku.edu}
\affiliation{Department of Physics, Sungkyunkwan University, Suwon 16419, Korea}
\date{\today}

\begin{abstract} We propose a simple procedure by which the interaction parameters of the classical spin Hamiltonian can be determined from the knowledge of four-point correlation functions and specific heat. The proposal is demonstrated by using the correlation and specific heat data generated by Monte Carlo method on one- and two-dimensional Ising-like models, and on the two-dimensional Heisenberg model with Dzyaloshinkii-Moriya interaction. A recipe for applying our scheme to experimental images of magnetization such as those made by magnetic force microscopy is outlined.
\end{abstract}
\maketitle

Condensed matter, either in their natural form or as synthesized in laboratories, are invariably complicated and have complex interactions among its constituents. Spin-spin interaction in magnetic insulators for one thing can have ranges well beyond the first neighbor, and yet models almost uniquely focus on cases with only one or just a few interaction parameters. Even between a pair of nearest-neighbor spins, the interactions can take on symmetric or anti-symmetric forms, either preserving the spin rotation symmetry or breaking it altogether. Deducing the proper interaction parameters is also of paramount importance in the search for exotic spin liquid such as the Kitaev spin liquid~\cite{kitaev}.

It has been a rich and fruitful practice in physics to rely on insights and quasi-exact results from simplest models to interpret the phenomena arising in complex materials, knowing that the actual interaction Hamiltonian generally bears more complexity than those simple models suggest. With the synthesis of materials of ever-increasing complexity and novelty, a corresponding improvement in the technique to identify the microscopic interaction parameters must find parallel advances. One powerful technique to identify spin-spin interaction parameters in magnets is the fit made to the inelastic neutron scattering data by the spin-wave spectrum worked out from model Hamiltonians. We hereby propose a simple scheme that could accomplish similar task. The idea is illustrated with Ising-type spin models in one and two dimensions, and a Heisenberg-type spin model in two dimensions.

Suppose we had a one-dimensional Hamiltonian

\ba H = - \sum_{1 \le i \le L} \left( \sum_{1 \le r \le R} J_{r} \sigma_i \sigma_{i+r} \right) \label{eq:1D-H} \ea
made up of Ising variables $\sigma_i = \pm 1$ at the site $i$ in one-dimensional lattice of length $L$, and the interaction $J_{r}$ extends up to $R$-th neighbors. The simplest case $J_1 = 1$ and $J_{r \neq 1} =0$ is the one-dimensional Ising model. Translational invariance is assumed in this class of models. Statistical properties of the model are easy to generate by means of the Monte Carlo (MC) simulation. Given some material whose interactions are assumed to fit the above $H$ with some choice of $J_r$'s, and some of its thermodynamic properties known experimentally, would it be possible to fix the parameters $J_r$ by virtue of the known experimental input? We claim the answer is in the affirmative, specifically if the four-point correlation function and the specific heat are known accurately as a function of temperature. We support our claim and illustrate the recovery procedure using the statistical data generated by the MC simulation.

A well-known theorem of equilibrium statistical mechanics is $(k_B = 1)$

\ba \langle H^2 \rangle - \langle H \rangle^2 = T^2 C(T) \ea
where $C(T)$ refers to the specific heat. In terms of the general Hamiltonian (\ref{eq:1D-H}), one can re-cast the identity as

\ba & \sum_{r,r'} J_r {\cal C}_{r,r'} J_{r'} = T^2 C(T), \nn
& {\cal C}_{r,r'} = \langle \Sigma_r \Sigma_{r'} \rangle -  \langle \Sigma_r \rangle \langle \Sigma_{r'} \rangle, \ea
where $\Sigma_r = \sum_i \sigma_i \sigma_{i+r}$. The four-point correlation functions $C_{r,r'}$ form a temperature-dependent, real and symmetric matrix. One can write the identity in the matrix form

\ba {\cal J}^T {\cal C}(T) {\cal J} = T^2 C(T) , \ea
where ${\cal J}$ is a vector consisting of all the interaction parameters.

With the given knowledge of ${\cal C}(T)$ and $C(T)$ over a sufficiently wide temperatue range, it becomes a matter of determining ${\cal J}$ that best reproduces the thermodynamic identity. Defining the difference function $D({\cal J}, T) = {\cal J}^T {\cal C}(T) {\cal J} - T^2 C(T)$, the cost function to minimize is

\ba I[{\cal J}] = \sum_T [D({\cal J}, T)]^2 . \ea
The summation $\sum_T$ takes place over all temperatures for which correlation and specific heat data are available. Applying the gradient descent (GD)

\ba {\partial I \over \partial J_r } = 4 \sum_T D (T) \Bigl( \sum_{r'} {\cal C}_{r,r'} (T) J_{r'} \Bigr) ,  \label{eq:gradient} \ea
one can update the parameters $J_r$ iteratively until convergence is reached, thus completing the ``statistical recovery" of the original Hamiltonian.

\begin{figure}
  \centering
\includegraphics[scale=0.45]{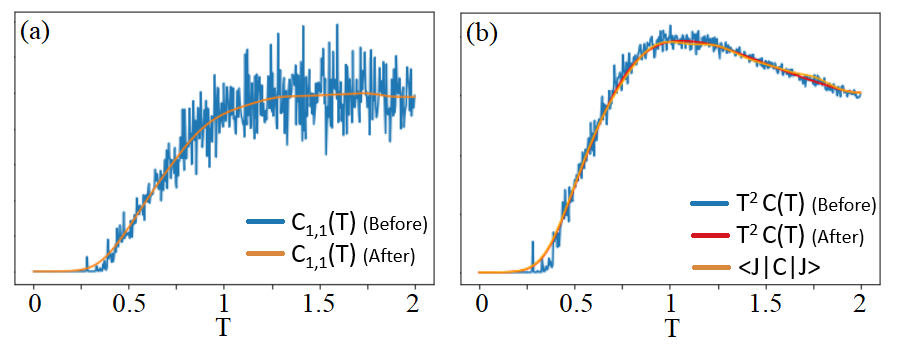}
\caption{(a) Correlation function ${\cal C}_{1,1}(T)$ of the one-dimensional Ising-like model (\ref{eq:1D-H}) with$(J_1, J_2, J_3 ) = (1, 0.5 , 0.33)$, before and after smoothing. (b) $T^2 C(T)$ before and after smoothing. The ${\cal J}^T {\cal C}(T) {\cal J}$ curve shown here using the recovered parameters $(J_1, J_2, J_3 ) = (1.017, 0.506, 0.329)$ is indistinguishable from the smoothened $T^2 C(T)$ curve.}\label{fig:1}
\end{figure}

In practice, some pre-conditioning of the data is required to ensure the convergence of the GD scheme. The inevitable noise from statistical fluctuations in the temperature dependence of the correlation functions ${\cal C}_{r,r'} (T)$ as well as the specific heat function $C(T)$  carries over to the gradient $\partial I/\partial J_r$ in (\ref{eq:gradient}), creating unwanted local minima in the cost function's landscape. On the other hand, the GD method proved to work very well if we first smooth out both functions with a Gaussian filter, and then apply the descent scheme. The correlation function ${\cal C}_{1,1}(T)$ and the specific heat $C(T)$ before and after smoothing are shown in Fig. \ref{fig:1}.

Another recipe we found crucial in the successful implementation of the GD method is rooted on the physically motivated interaction hierarchy $|J_r | > |J_{r'}|$ when $r$ is less than $r'$.
Instead of updating all the parameters at once, we first update $J_1$ using the gradient $\partial I/\partial J_1$, keeping all other $J_r$'s fixed. After, say, 1000 iterations for $J_1$, we start updating $J_2$ according to $\partial I/\partial J_2$ while keeping all $J_{r\neq 2}$ fixed. Once the update reaches the final $J_r$, we come back to $J_1$ and start over the iteration. A small enough cost function is achieved after repeating this procedure $\sim 10^3$ times.

The validity of our scheme was tested for one-dimensional ferromagnetic model (\ref{eq:1D-H}) of length $L=100$ with $(J_1, J_2, J_3) = (1, 0.5, 0.33)$. MC annealing was used to generate the correlation matrix ${\cal C}_{r,r'}(T)$ and the specific heat $C(T)$ over $0< T< 2$, and the GD scheme was applied in the prescribed manner. Five hundred temperature steps were taken. After the GD iteration is complete, we find the recovery parameters $(J_1, J_2, J_3 ) = (1.017, 0.506, 0.329)$ in close proximity to the original values, irrespective of the initial parameters chosen for the iteration.

%

\begin{figure}
  \centering
\includegraphics[scale=0.5]{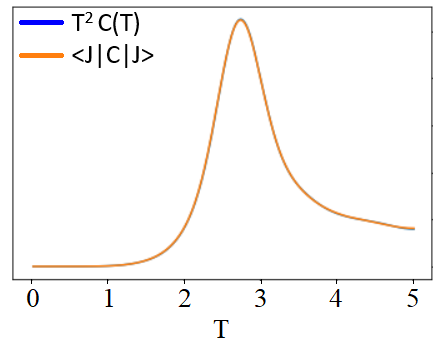}
\caption{Plots of $T^2 C(T)$ and ${\cal J}^T {\cal C}(T) {\cal J}$ for two-dimensional Ising-type model with interactions $(J_1, J_2, J_3) = (1.0, 0.7, 0.5)$ on $20\times20$ lattice.  Two curves are indistinguishable.}\label{fig:2}
\end{figure}

The scheme is subsequently applied to two-dimensional ferromagnetic Ising-type model with first- to third-neighbor interactions, $(J_1, J_2 , J_3 ) = (1, 0.7, 0.5)$, on the $L\times L$ square lattice. Figure \ref{fig:2} shows $T^2 C(T)$ both from original MC annealing and from the statistical recovery procedure on $20\times 20$ lattice. One sees only one curve because of the heavy overlap of the original and recovered plots. The interaction parameters obtained through the statistical recovery were $(J_1, J_2 , J_3 ) = (0.992, 0.691, 0.507)$ after 6000 sweeps through the parameters.

Deducing parameters of the Heisenberg-type spin Hamiltonian through our recovery procedure is a greater challenge. We consider as an example the Heisenberg-Dzyaloshinskii-Moriya-Zeeman (HDMZ) Hamiltonian given by

\ba H_{\rm HDMZ} & = & -J \sum_i \v S_i \cdot ( \v S_{i+\hat{x} } + \v S_{i+\hat{y} } ) \nn
& + & D \sum_i \v S_i \cdot (\v S_{i+\hat{x} } \times \hat{y} - \v S_{i+\hat{y} } \times \hat{x} ) \nn
& - & \v B \cdot \sum_i \v S_i . \label{eq:HDMZ} \ea
Its properties and phase diagram are well-known~\cite{han-book}. Although this model has been primarily used to understand the properties of skyrmions~\cite{han-book,nagaosa-review}, we adopt this model here for the sake of illustrating the statistical recovery procedure.

Suppose now that we did not know the exact structure of the microscopic Hamiltonian, and instead had to assume the more general spin-spin interaction

\ba H = \sum_{i} \Bigl[  \left( \sum_{a=x,y} \sum_{\alpha,\beta=x,y,z } J_a^{\alpha\beta} S_i^\alpha S_{i+\hat{a}}^\beta \right)  - B S_i^z  \Bigr] . \label{eq:fitting-H} \ea
In the most general circumstance we have a total of 18 fitting parameters $J_a^{\alpha\beta}$. The energy variance for $B=0$ follows from

\ba \langle H^2 \rangle - \langle H \rangle^2 & = & \sum_{a,a'} \sum_{\alpha\beta, \alpha'\beta'} J^{\alpha\beta}_a J^{\alpha'\beta'}_{a'}  {\cal C}_{aa'}^{\alpha\beta, \alpha' \beta'} \nn
{\cal C}_{aa'}^{\alpha\beta, \alpha' \beta'}  & = & \langle \Sigma_a^{\alpha\beta} \Sigma_{a'}^{\alpha' \beta'} \rangle - \langle \Sigma_a^{\alpha\beta}  \rangle \langle \Sigma_a^{\alpha' \beta' } \rangle , \ea
where $\Sigma_a^{\alpha\beta} = \sum_i S_i^\alpha S_{i+\hat{a}}^\beta$. Much more complicated variance as well as the GD formula have to be worked out for $B \neq 0$, which only adds complication to the recovery scheme. As far as the proof-of-concept demonstration goes, we find it sufficient to focus on $B=0$.

\begin{figure}
  \centering
\includegraphics[scale=0.5]{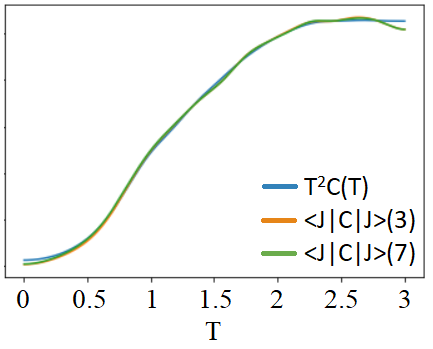}
\caption{Plots of $T^2 C(T)$ and ${\cal J}^T {\cal C}(T) {\cal J}$ for HDMZ model (\ref{eq:HDMZ}) with $B=0$ and $(J,D)=(1,\sqrt{6})$. Both 3 and 7 parameter fits were used with very similar results. }\label{fig:3}
\end{figure}

Taking a $18\times18$ lattice with $D=\sqrt{6}$ and $J=1$ corresponding to the spiral period of six lattice constants, we generated the correlation matrix and the specific heat function over $0<T < 3$ and used the GD scheme to reproduce the fitting parameters in (\ref{eq:fitting-H}). Since all interactions are nearest-neighbor, the parameters were updated simultaneously. First we impose a restriction that all diagonal interactions are equal, $J_a^{xx} = J_a^{yy} = J_a^{zz} = -J$ $(a=x,y)$, and that the only off-diagonal interactions are $J_{x}^{zx} = -J_x^{xz}$ and $J_{y}^{yz} = -J_y^{zy}$. In this 3-parameter fitting scheme we recover $(J, J_x^{zx}, J_y^{zy} ) = (0.999, 2.473, 2.423)$, in excellent agreement with the original $(J, J_x^{zx}, J_y^{zy} ) =(1,\sqrt{6},\sqrt{6})$. Relaxing the conditions slightly so that $J_a^{\alpha\beta}=-J_a^{\beta\alpha}$ for $\alpha\neq \beta$, and $J_a^{\alpha\alpha}=-J$, we obtain the seven-parameter fit with $(J, J_x^{zx}, J_y^{zy}, J^{yx}_y , J^{zy}_x, J^{xy}_x , J^{xz}_y ) = (0.924, 2.396, 2.572, 0.289, 0.239, 0.054, 0.254)$. The leakage into the parameters $(J^{yx}_y , J^{zy}_x, J^{xy}_x , J^{xz}_y )$ that did not exist in the original Hamiltonian is an unavoidable occurrence in the GD optimization; the more parameters are involved, the better becomes the fit. Parameters which ought to be zero, or equal to each other by symmetry, are better set as such in the GD iteration. Otherwise the GD iteration will choose to break such constraint in search of ever-improving fit to the target function. Figure \ref{fig:3} shows the recovered $T^2 C(T)$ in excellent agreement with the original curve.

In order to implement the recovery scheme proposed here in the actual experiment, we need information not only of the specific heat, but also of the four-point correlation functions. The best chance of obtaining this information comes from surface-sensitive measurement of the local magnetization. Examples are spin-polarized scanning tunneling microscopy (SPSTM)~\cite{SPSTM}, magnetic force microscopy (MFM)~\cite{MFM}, and Lorentz transmission electron microscopy (LTEM)~\cite{LTEM}, all of which are being actively used in the investigation of low-dimensional magnets. Measuring the specific heat of a truly two-dimensional material poses an obvious challenge, but there is progress in recent years to measure the thermodynamic quantity of single and multi-layer graphene~\cite{2D-specific-heat}. In layered materials with very weak inter-layer interaction, the measured bulk specific heat can be translated into the per-layer quantity, while surface probes reveal the four-point correlations of the magnetic moment within the plane.

We outline a prescription, partly described in an earlier publication~\cite{kumar18}, to extract four-point correlations from the surface data. Let's say we are given the $512\times 512$ pixel image of an MFM measurement on some surface where each pixel represents the local magnetization normal to the plane. The image can be cut into $16\times16$ pieces of equal sizes, each piece containing $32\times32$ pixels. These 256 pieces cut out from one large $512\times512$ batch constitute the {\it ensemble} of states corresponding to the same external conditions such as temperature and magnetic field. Taking $M=256$ as the number of states, the ensemble average procedure goes as

\ba && \langle \sum_{i,j}  \sigma_i \sigma_{i+r} \sigma_j \sigma_{j+r'}  \rangle - \langle \sum_i \sigma_i \sigma_{i+r} \rangle \langle \sum_j \sigma_j \sigma_{j+r'} \rangle\nn
&& = M^{-1} \sum_{c=1}^M \sum_{i,j}  \sigma_i^{(c)} \sigma_{i+r}^{(c)}\sigma_j^{(c)} \sigma_{j+r'}^{(c)} \nn
&& ~~~~~ - M^{-2} \Bigl( \sum_{c, i } \sigma_i^{(c)} \sigma_{i+r}^{(c)} \Bigr) \Bigl( \sum_{c,j} \sigma_j^{(c)} \sigma_{j+r'}^{(c)} \Bigr) . \ea
The summation $\sum_i$ and $\sum_j$ as well as the positions $r$ and $r'$ are confined within the $32\times 32$ pixel area. This is not a severe restriction in practice since the interaction parameters $J_r$ are expected to die off quickly with separation $r$.

Oftentimes the resolution of the images is not truly atomic scale, as with the MFM measurement. Each pixel in the MFM image represents an average of the local magnetization within the resolution window, much like the coarse-graining process in the real-space renormalization group theory. In that case the interaction Hamiltonian deduced by our procedure would be the coarse-grained version of the true microscopic Hamiltonian. Even a microscopic Hamiltonian involving only the nearest-neighbor interaction is known to generate longer-ranged interactions upon coarse-graining~\cite{stat-mech}, and our demonstration of the fitting procedure in terms of several interaction parameters is of practical relevance.

An analogous proposal was made in Ref. \onlinecite{qi17} for the quantum case, which argued that a single wave function and the four-point correlations obtained with respect to it is sufficient to recover the parameters of the original microscopic Hamiltonian. In detail, the procedure proposed in Ref. \onlinecite{qi17} is quite different from ours, and assumes the full knowledge of either the wave function or its four-point correlation functions, both of which are extremely challenging to obtain experimentally. Our proposal is based on simple application of classical statistical mechanics, and assumes knowledge of the ensemble average rather than the quantum expectation value. An enormous range of Ising-like magnets have been identified and thoroughly studied in the past~\cite{Ising-magnets}, and we believe direct application of our scheme to such magnets should be feasible. Fits to the specific heat and the magnetic susceptibility as a means to deduce interaction parameters of the Ising-like magnet have persisted over the years~\cite{Ising-magnets}. Inelastic neutron scattering also offers a strong venue for determining the interaction parameters in insulating magnets. One advantage of our method over existing ones comes from the implementation of the GD scheme, which automatically finds the appropriate set of parameters once the four-point correlation and the specific heat are known with sufficient accuracy. No fine-tuning of the parameters by hand is required, nor is it possible.

This work was supported by Samsung Science and Technology Foundation under Project Number SSTF-BA1701-07.


\end{document}